\documentclass[twocolumn,showpacs,preprintnumbers,pra]{revtex4}
\usepackage{amsmath}
\usepackage{dcolumn}
\usepackage{bm}
\usepackage{graphicx}
\usepackage{amsfonts}
\usepackage{amssymb}
\usepackage{bbm}
\usepackage{float}
\usepackage[dvipdfm]{hyperref}

\begin{document}

\title{Quantum-speed-limit time for multiqubit open systems}
\author{Chen Liu}
\email{tarksar@sina.com}
\author{Zhen-Yu Xu}
\altaffiliation{}
\email{zhenyuxu@suda.edu.cn}
\author{Shiqun Zhu}
\altaffiliation{}
\email{szhu@suda.edu.cn}
\affiliation{College of Physics, Optoelectronics and Energy, Soochow University, Suzhou
215006, China}

\begin{abstract}
Quantum-speed-limit (QSL) time captures the intrinsic minimal time interval
for a quantum system evolving from an initial state to a target state. In
single qubit open systems, it was found that the memory (non-Markovian)
effect of environment plays an essential role in shortening QSL time or,
say, increasing the capacity for potential speedup. In this paper, we
investigate the QSL time for multiqubit open systems. We find that for a
certain class of states the memory effect still acts as the indispensable
requirement for cutting the QSL time down, while for another class of states
this takes place even when the environment is of no memory. In particular,
when the initial state is in product state $\left\vert 111\cdots
1\right\rangle $,\ there exists a sudden transition from no capacity for
potential speedup to potential speedup in a memoryless environment. In
addition, we also display evidence for the subtle connection between QSL
time and entanglement that weak entanglement may shorten QSL time even more.
\end{abstract}

\maketitle

\section{Introduction}

Quantum-speed-limit (QSL) time \cite{MT,ML}, the intrinsic minimal time
interval for a quantum system evolving from an initial state to a target
state, is of crucial importance in the fields of quantum computation \cite%
{quantum computer}, quantum control \cite%
{control1,control2,control3,control4,control5}, quantum metrology \cite%
{metrology,metrology1,metrology2}, and non-equilibrium thermodynamics \cite%
{thermo}. Recent decades have witnessed a great deal of research on QSL time
both in closed \cite{c1,c2,c3,c4,c5,c6,c7,c8,c9,c10,c11,c12,new1} and open
systems \cite{xopen0,xopen1,QSLopen1,QSLopen2,QSLopen3,new2,Xu1,Xu2}. In
particular, a QSL time based on the Schatten $p$ norm for an arbitrarily
driven open system was presented by Deffner and Lutz \cite{QSLopen3}

\begin{equation}
\tau _{QSL}=\frac{\sin ^{2}[B(\rho ,\rho _{\tau })]}{\min \left\{
E_{1}^{\tau },E_{2}^{\tau },E_{\infty }^{\tau }\right\} },  \label{QSL}
\end{equation}%
where $B(\rho ,\rho _{\tau })=\arccos (\sqrt{\langle \phi |\rho _{\tau
}|\phi \rangle })$ denotes the Bures angle between the initial state $\rho
=\left\vert \phi \right\rangle \langle \phi |$ and the target state $\rho
_{\tau }$, which is governed by the time-dependent master equation $\dot{\rho%
}_{t}=L_{t}\rho _{t}$ ($L_{t}$ is a superoperator). It is reasonable to
employ the Bures angle above as a measure for distance between an pure state
and a mixed state, since the important Riemannian feature is satisfied under
such a circumstance \cite{QSLopen1}. The denominator $E_{p}^{\tau }$ in Eq. (%
\ref{QSL}) represents the average of $||L_{t}\rho _{t}||_{p}$ over actual
driving time duration $\tau $, i.e., $E_{p}^{\tau }=(1/\tau )\int_{0}^{\tau
}||L_{t}\rho _{t}||_{p}dt$. One important application of the QSL time is to
evaluate the speed of quantum evolutions under the following two scenarios:

(I). The fastest evolution appears when the actual driving time $\tau $
achieves the QSL time $\tau _{QSL}$, i.e., $\tau /\tau _{QSL}=1.$ For slower
evolutions, $\tau /\tau _{QSL}>1$ and the slower the evolution, the higher
ratio $\tau /\tau _{QSL}$ should be. This point of view is clear and
unambiguous and has been widely utilized in the field of entanglement
assisted speedup of quantum evolutions \cite{c5,en2,en3,en4}.

(II). From another point of view, $\tau /\tau _{QSL}=1$ indicates the
evolution is already the fatest, and possesses no potential capacity for
further acceleration, while for $\tau /\tau _{QSL}>1,$ the higher ratio $%
\tau /\tau _{QSL}$ (or equivalently, the much shorter $\tau _{QSL}$), the
greater the capacity for potential speedup will be. This viewpoint has been
adopted in exploring the memory effect, characterized by non-Markovianity
\cite{nonM-review1,nonM-review2}, on the speed of quantum evolution \cite%
{QSLopen3,Xu1,Xu2}. It is found that in the damped Jaynes-Cummings model of
a single qubit, the transition from no potential capacity of speedup ($\tau
/\tau _{QSL}=1$) to potential speedup ($\tau /\tau _{QSL}>1$), or say a
reduction of QSL in quantum evolution is just the critical point when the
memoryless environment becomes of memory \cite{QSLopen3,Xu1}. Furthermore,
if evolution of a\ qubit is already along the QSL time which means its
fastest speed in nature environment, the study on it will probably not so
urgent as the case of $\tau /\tau _{QSL}>1$ which have speedup potential.

Although the memory effect of environment plays a decisive role in the
potential acceleration of quantum evolution in a single qubit case, the
question may arise whether it will still be true even in multi-qubit cases.
This question is of particular interest, for the QSL time of multi-qubit
open systems has now caught increasing attention and several interesting
phenomena were discovered. For instances, with the QSL time defined in Ref. (%
\cite{QSLopen1}), Taddei \textit{et al}. illustrated that the separable
states of a multi-qubit system under Markovian dephasing channels perform
the same speedup of quantum evolution as the entangled states when the
number of qubits is large enough. On the other hand, del Campo \textit{et al.%
} demonstrated, with the bound introduced in Ref. (\cite{QSLopen2}), that
multi-qubit Greenberger-Horne-Zeilinger (GHZ) states and separable states
under Markovian and non-Markovian dephasing are equivalent in metrological
parameter estimation \cite{QSLopen2}. In this paper, mainly focused on
Scenario 2, we show that for a class of multiqubit states memory effect is a
requisite for raising potential ability of speeding up quantum evolution,
while for another class of multi-qubit states, memoryless environment is
ready to realize the above result. Especially when the open system is
initially prepared in product state $\left\vert 111\cdots 1\right\rangle ,$
a transition from no potential capacity for speedup to possess speedup
ability is achieved.

The paper is organized as follows: In Sec. II, we present the QSL time for
typical two-qubit, three-qubit, and n-qubit states, respectively. Discussion
on the role of entanglement in QSL time is performed in Sec. III. Finally,
conclusions are drawn in Sec. IV.

\section{Quantum limits to multi-qubit dynamical evolution}

We consider N independent two-level atoms (open system) each locally
coupling to a leaky vacuum cavity (environment). The dynamics of the
multi-qubit open system is fully determined by each pair of atom cavities
\cite{Bellomo} with the following Hamiltonian \cite{Book Open}

\begin{equation}
H=\omega _{0}\sigma _{+}\sigma _{-}+\sum_{k}\omega _{k}a_{k}a_{k}^{\dag
}+i\sum_{k}g_{k}(a_{k}^{\dag }\sigma _{-}-a_{k}\sigma _{+}),
\label{Hamiltonian}
\end{equation}%
where $\omega _{0}$ is the resonant transition frequency of the atom between
the excited state $|1\rangle $ and the ground state $|0\rangle $, and $%
\sigma _{\pm }$ are the Pauli raising and lowering operators. $\omega _{k}$
and $a_{k}(a_{k}^{\dag })$ denote the frequency and the
annihilation(creation) operators of the $k$th mode of the cavity with $g_{k}$
the corresponding real coupling constant. The master equation for the
reduced density matrix of the atom in the Schr\"{o}dinger picture is given
by $\dot{\rho}_{t}=L_{t}\rho _{t}$ with
\begin{equation}
L_{t}\rho _{t}=i\delta _{t}\left[ {\sigma }_{+}{\sigma }_{-},\rho _{t}\right]
+\gamma _{t}\left( {\sigma }_{+}{\sigma }_{-}\rho _{t}+\rho _{t}{\sigma }_{+}%
{\sigma }_{-}-2{\sigma }_{-}\rho _{t}{\sigma }_{+}\right) ,
\label{master equation}
\end{equation}%
where $\delta _{t}$=Im$(\dot{c}_{t}/c_{t})$ and $\gamma _{t}$=Re$(\dot{c}%
_{t}/c_{t})$ are the time-dependent Lamb shift and decay rate respectively,
and $c_{t}$ is the decoherence function relying on the particular structure
of cavity reservoirs \cite{Book Open}. The reduced density matrix of the
atom with an initial state $\rho =(\rho _{mn})$ takes the form

\begin{equation}
\rho _{t}=\left[
\begin{array}{cc}
\rho _{11}|c_{t}|^{2} & \rho _{10}c_{t} \\
\rho _{01}c_{t}^{\ast } & 1-\rho _{11}|c_{t}|^{2}%
\end{array}%
\right] ,  \label{density matrix}
\end{equation}%
where the excited state population $|c_{t}|^{2}$ is denoted by $P_{t}$ in
the following.

\subsection{Two-qubit cases}

As the exact form of QSL time for a general pure initial state is
cumbersome, we only consider two typical Bell-type initial states
respectively, i.e., $\left\vert \Psi _{1}\right\rangle =\alpha |01\rangle +%
\sqrt{1-\alpha ^{2}}|10\rangle $ and $\left\vert \Psi _{2}\right\rangle
=\alpha |11\rangle +\sqrt{1-\alpha ^{2}}|00\rangle $ with $\alpha \in
\lbrack 0,1]$. According to the definition of QSL time in Eq. (\ref{QSL}),
we have
\begin{equation}
\frac{\tau }{\tau _{QSL}}=\frac{\int_{0}^{\tau }\left\vert \overset{\cdot }{P%
}_{t}\right\vert dt}{1-P_{\tau }}  \label{QSL2-1}
\end{equation}%
for state $\left\vert \Psi _{1}\right\rangle $ and

\begin{equation}
\frac{\tau }{\tau _{QSL}}=\frac{\int_{0}^{\tau }\max \left\{ \left\vert
\overset{\cdot }{P}_{t}(2\alpha P_{t}-\alpha \pm 1)\right\vert \right\} dt}{%
\alpha (1-P_{\tau }^{2})}  \label{QSL2-2}
\end{equation}%
for state $\left\vert \Psi _{2}\right\rangle $, where $\overset{\cdot }{P}%
_{t}=\partial _{t}P_{t}.$

\begin{figure}[tbp]
\centering{}\includegraphics[width=3.5in]{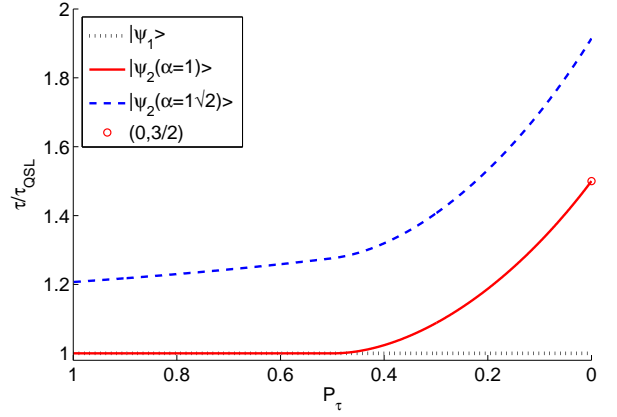}
\caption{(Color online) Quantum-speed-limit time ratio $\protect\tau _{QSL}/%
\protect\tau $ of a two-qubit open system under a memoryless environment as
a function of $P_{\protect\tau }$ with initial states $\left\vert \Psi
_{1}\right\rangle =\protect\alpha |01\rangle +\protect\sqrt{1-\protect\alpha %
^{2}}|10\rangle $ (black dotted line), $\left\vert \Psi _{2}(\protect\alpha %
=1)\right\rangle =|11\rangle $ (red solid curve), and $\left\vert \Psi _{2}(%
\protect\alpha =1/\protect\sqrt{2})\right\rangle =$ $(|11\rangle +|00\rangle
)/\protect\sqrt{2}$ (blue dashed curve). The red circle indicates the
maximal speedup of quantum evolution for state $\left\vert \Psi _{2}(\protect%
\alpha =1)\right\rangle =|11\rangle .$}
\end{figure}

Obviously, the QSL time ratio $\tau _{QSL}$ of state $\left\vert \Psi
_{1}\right\rangle $ is the same as the exact form in single qubit cases
discovered in Refs \cite{QSLopen3,Xu1}, where the reduction in QSL time ($%
\tau /\tau _{QSL}>1$) only occurs when the environment is of memory;
otherwise the evolution will always be along the QSL time ($\tau /\tau
_{QSL}\equiv 1$). In order to illustrate this phenomenon clearly, the QSL
time ratio $\tau /\tau _{QSL}$ of initial state $\left\vert \Psi
_{1}\right\rangle $ (black dotted line) is depicted in Fig. 1 under a
memoryless environment, i.e., $\left\vert \overset{\cdot }{P}_{t}\right\vert
=-\overset{\cdot }{P}_{t}$ with the population $P_{t}$ monotonically
decreasing from 1 to the target $P_{\tau }$.

However, a complex but interesting phenomenon appears for initial state $%
\left\vert \Psi _{2}\right\rangle $: QSL reduction can also take place even
when the environment is of no memory. For instances, $\tau /\tau _{QSL}$
versus $P_{\tau }$ of initial states $\left\vert \Psi _{2}(\alpha
=1)\right\rangle =|11\rangle $ (red solid curve) and $\left\vert \Psi
_{2}(\alpha =1/\sqrt{2})\right\rangle =$ $(|11\rangle +|00\rangle )/\sqrt{2}$
(blue dashed curve) are depicted, respectively, in Fig. 1, where $\tau /\tau
_{QSL}\geq 1$ clearly illustrates the intrinsic acceleration potential of
quantum evolution under a memoryless environment. Especially, there exists a
sudden change point in QSL time with the critical point $P_{\tau }=1/2$ for
initial state $\left\vert \Psi _{2}(\alpha =1)\right\rangle =$ $|11\rangle $%
. To explain this phenomenon, we trace back to $||L_{t}\rho _{t}||_{\infty }$
defined in Eq. (\ref{QSL}) with the following expression

\begin{equation}
||L_{t}\rho _{t}||_{\infty }=\left\{
\begin{array}{l}
-\overset{\cdot }{P}_{t}(2\alpha ^{2}P_{t}-\alpha ^{2}+\alpha ), \\
\overset{\cdot }{P}_{t}(2\alpha ^{2}P_{t}-\alpha ^{2}-\alpha ),%
\end{array}%
\right.
\begin{array}{c}
P_{t}\geqslant \frac{1}{2}, \\
P_{t}<\frac{1}{2},%
\end{array}
\label{QSL2-3}
\end{equation}%
where we have employed the condition of memoryless environment $\left\vert
\overset{\cdot }{P}_{t}\right\vert =-\overset{\cdot }{P}.$ Therefore, the
QSL time ratio $\tau /\tau _{QSL}$ of Eq. (\ref{QSL2-2}) can be conveniently
calculated as
\begin{equation}
\frac{\tau }{\tau _{QSL}}=\left\{
\begin{array}{ll}
\frac{1+\alpha P_{\tau }}{\alpha (1+P_{\tau })}, & P_{\tau }\geqslant \frac{1%
}{2}, \\
\frac{2(1-P_{\tau })(1-\alpha P_{\tau })+\alpha }{2\alpha (1-P_{\tau }^{2})}
& P_{\tau }<\frac{1}{2}.%
\end{array}%
\right.  \label{QSL2-4}
\end{equation}%
One may check that $\tau /\tau _{QSL}\geq 1$ is always satisfied. In
particular, when the initial state is in $\Psi _{2}(\alpha =1)=|11\rangle $,
the QSL time ratio yields
\begin{equation}
\frac{\tau }{\tau _{QSL}}=\left\{
\begin{array}{ll}
1, & P_{\tau }\geqslant \frac{1}{2}, \\
\frac{2(1-P_{\tau })^{2}+1}{2(1-P_{\tau }^{2})} & P_{\tau }<\frac{1}{2}.%
\end{array}%
\right.  \label{QSL2-5}
\end{equation}%
The sudden change point of QSL time is therefore justified.

\subsection{Three-qubit cases}

In this subsection, we also consider two typical three-qubit states, i.e., W
type state $\left\vert \Psi _{3}\right\rangle =\alpha |001\rangle +\beta
|010\rangle +\sqrt{1-\alpha ^{2}-\beta ^{2}}|100\rangle $ and GHZ type state
$\left\vert \Psi _{4}\right\rangle =\alpha |111\rangle +\sqrt{1-\alpha ^{2}}%
|000\rangle .$ According to Eq. (\ref{QSL}), the expressions of the QSL time
ratio are obtained:
\begin{equation}
\frac{\tau }{\tau _{QSL}}=\frac{\int_{0}^{\tau }\left\vert \overset{\cdot }{P%
}_{t}\right\vert dt}{1-P_{\tau }}  \label{QSL3-1}
\end{equation}%
for state $\left\vert \Psi _{3}\right\rangle $ and

\begin{equation}
\frac{\tau }{\tau _{QSL}}=\frac{\int_{0}^{\tau }\max \left\{ \left\vert
\overset{\cdot }{P}_{t}(\pm \frac{3}{2}X+3\alpha P_{t}-\frac{3}{2}\alpha
)\right\vert \right\} dt}{\alpha +\alpha (1-\alpha ^{2})P_{\tau }(3-2P_{\tau
}^{\frac{1}{2}}-3P_{\tau })+\alpha (2\alpha ^{2}-1)P_{\tau }^{3}},
\label{QSL3-3}
\end{equation}%
for state $\left\vert \Psi _{4}\right\rangle $, where $X=(4\alpha
^{2}P_{t}^{4}-8\alpha ^{2}P_{t}^{3}+8\alpha ^{2}P_{t}^{2}-5\alpha
^{2}P_{t}+P_{t}+\alpha ^{2})^{\frac{1}{2}}.$

Equation (\ref{QSL3-1}) bears a resemblance to the case of state\ $%
\left\vert \Psi _{1}\right\rangle $, implying that the reduction in QSL only
occurs in a memory environment. As for Eq. (\ref{QSL3-3}), we consider a
special case, i.e., $\left\vert \Psi _{4}(\alpha =1)\right\rangle
=|111\rangle $ under the environment of no memory ($\left\vert \overset{%
\cdot }{P}_{t}\right\vert =-\overset{\cdot }{P}$). Therefore, the Eq. (\ref%
{QSL3-3}) can be simplified as:

\begin{equation}
\frac{\tau }{\tau _{QSL}}=\left\{
\begin{array}{ll}
1, & P_{\tau }\geqslant \frac{1}{2}, \\
\frac{(-3P_{\tau }+3P_{\tau }^{2}-P_{\tau }^{3}+\frac{7}{4})}{1-P_{\tau }^{3}%
} & P_{\tau }<\frac{1}{2},%
\end{array}%
\right.  \label{QSL3-5}
\end{equation}%
indicating that there also exists a sudden transition of speedup potential
of quantum evolution even the environment is of no memory.

\subsection{N-qubit cases}

In this subsection, we show that the above phenomena are ubiquitous in
n-qubit cases (n is an arbitrary positive integer). It is easy to check that
if the n-qubit open system is initially prepared in state $\alpha
_{1}\left\vert 100\cdots 0\right\rangle +\alpha _{2}\left\vert 010\cdots
0\right\rangle +\cdots +\alpha _{N}\left\vert 000\cdots 1\right\rangle ,$
with $\sum_{j=1}^{N}\alpha _{j}^{2}=1,$ the QSL ratio is exactly the same as
Eqs. (\ref{QSL2-1}) and (\ref{QSL3-1}). Therefore, the memory effect of
environment becomes the essential condition for the speedup potential emerge
in quantum evolution.

However, if the initial state is in $\left\vert 11\cdots 1\right\rangle $,
the QSL ratio is given by

\begin{equation}
\frac{\tau }{\tau _{QSL}}=\left\{
\begin{array}{ll}
1, & P_{\tau }\geqslant \frac{1}{2}, \\
\frac{\int_{0}^{\tau }\max \{|n\overset{\cdot }{P}_{t}P_{t}^{n-1}|,|-n%
\overset{\cdot }{P}_{t}(1-P_{t})^{n-1}|\}dt}{1-P_{\tau }^{n}} & P_{\tau }<%
\frac{1}{2}.%
\end{array}%
\right.  \label{QSLN-1}
\end{equation}%
In particular when the environment is memoryless, i.e., $\left\vert \overset{%
\cdot }{P}_{t}\right\vert =-\overset{\cdot }{P},$ Eq. (\ref{QSLN-1}) reduces
to
\begin{equation}
\frac{\tau }{\tau _{QSL}}=\left\{
\begin{array}{ll}
1, & P_{\tau }\geqslant \frac{1}{2}, \\
\frac{(1-P_{\tau })^{n}+1-(\frac{1}{2})^{n-1}}{1-P_{\tau }^{n}} & P_{\tau }<%
\frac{1}{2}.%
\end{array}%
\right.  \label{QSLN-2}
\end{equation}%
Clearly, $P_{\tau }=1/2$ is the critical point at which the open system
experiences the sudden change of QSL time under a memoryless environment.

Equation (\ref{QSLN-2}) also implies that there exists a maximal
acceleration potential condition for state $\left\vert 11\cdots
1\right\rangle $ when $P_{\tau }\rightarrow 0$, and the corresponding
minimal QSL time ratio is given by:

\begin{equation}
\frac{\tau }{\tau _{QSL}}|_{\max }=\frac{2^{n}-1}{2^{n-1}}.
\label{maximal speedup}
\end{equation}%
Especially when $n=2$, Eq. (\ref{maximal speedup}) grows to $3/2$, which is
marked in Fig. 1 as the red circle.

\subsection{Memory effect on QSL time}

\begin{figure}[tbp]
\centering{} \includegraphics[width=3.5in]{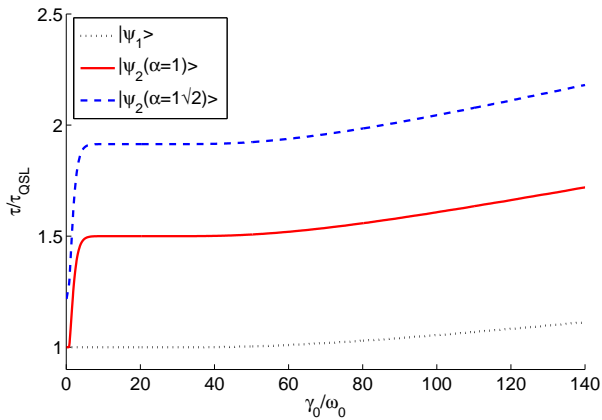}
\caption{(Color online) Quantum-speed-limit time $\protect\tau _{QSL}$ of a
Lorentzian spectral distribution model versus coupling strength $\protect%
\gamma _{0}.$ The black dotted, red solid, and blue dashed curves represent
the two-qubit initial states $\left\vert \Psi _{1}\right\rangle =\protect%
\alpha |01\rangle +\protect\sqrt{1-\protect\alpha ^{2}}|10\rangle $, $%
\left\vert \Psi _{2}(\protect\alpha =1)\right\rangle =|11\rangle $, and $%
\left\vert \Psi _{2}(\protect\alpha =1/\protect\sqrt{2})\right\rangle =$ $%
(|11\rangle +|00\rangle )/\protect\sqrt{2}$, respectively, with parameters $%
\protect\lambda =50$, $\protect\omega _{0}=1$, and $\protect\tau =1.$}
\end{figure}

In this subsection, we intend to show that memory effect of environment is
still an important element for quantum acceleration potential for
multi-qubit open systems. The memory environment we consider here is
characterized by the Lorentzian spectral distribution $J(\omega )=\frac{1}{%
2\pi }\frac{\gamma _{0}\lambda }{(\omega _{0}-\omega )^{2}+\lambda ^{2}}$ ,
where $\gamma _{0}$ is the Markovian decay rate and $\lambda $ is the
spectral width \cite{Book Open}. $P_{t}$ is now written as \cite{Book Open}%
\begin{equation}
P_{t}=e^{-\lambda t}\left\vert \cosh (\frac{dt}{2})+\frac{\lambda }{d}\sinh (%
\frac{dt}{2})\right\vert ^{2},  \label{damping}
\end{equation}%
where $d=\sqrt{2\gamma _{0}\lambda -\lambda ^{2}}$. In Fig. 2, we take
two-qubit initial states as an example and fix the driving time as $\tau =1.$%
The QSL time ratio $\tau /\tau _{QSL}$ versus coupling strength $\gamma
_{0}/\omega _{0}$ is plotted with three typical two-qubit states $\left\vert
\Psi _{1}\right\rangle $ (black dotted line), $\left\vert \Psi _{2}(\alpha
=1)\right\rangle $ (red solid curve), and $\left\vert \Psi _{2}(\alpha =1/%
\sqrt{2})\right\rangle $ (blue dashed curve), respectively, with $\lambda
=50 $, $\omega _{0}=1$, and we set $\tau _{QSL}=1$. According to Ref. (\cite%
{Book Open}), we know that $\gamma _{0}=\lambda /2$ is the transient point
from memoryless environment ($\gamma _{0}<\lambda /2$) to one of memory ($%
\gamma _{0}>\lambda /2$). As is clearly shown in Fig. 2, more capacity for
potential speedup will take place when the environment enters the memory
region.

\section{Discussion: Entanglement and QSL time}

Considering that the QSL time for multiqubit systems can be shortened in a
memoryless environment, it is natural to link this phenomenon to
entanglement, which only exists in multiqubit cases. In closed composite
systems, entanglement has been taken as a resource in the speedup of quantum
evolution [\cite{c5,en2,en3,en4}]. In this subsection, we go a step further
to the connection between entanglement and QSL time in bipartite open
systems. The initial state we consider here is an arbitrary pure state:

\begin{equation}
\left\vert \phi \right\rangle =\alpha _{1}|11\rangle +\alpha _{2}|10\rangle
+\alpha _{3}|01\rangle +\alpha _{4}|00\rangle ,  \label{arbitrary state}
\end{equation}%
with $\sum_{j=1}^{4}\alpha _{j}{}^{2}=1,$ which is generated by Monte Carlo
method, and the related entanglement is characterized by concurrence $C$ in
Ref. \cite{concurrence}, with $C=0$ for a disentangled state and $C=1$ for a
maximally entangled state.
\begin{figure}[tbp]
\centering{} \includegraphics[width=3.5in]{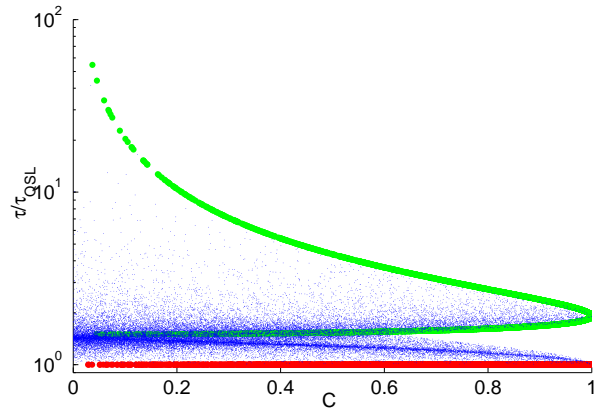}
\caption{(Color online) Quantum speed limit time ratio $\protect\tau _{QSL}/%
\protect\tau $ as a distribution of concurrence $C$ for 20000 randomly
generated initial pure states (tiny blue dots) under the damping model. The
dark red and light green dots represent QSL time ratio with states $%
\left\vert \Psi _{1}\right\rangle =\protect\alpha |01\rangle +\protect\sqrt{%
1-\protect\alpha ^{2}}|10\rangle $ and $\left\vert \Psi _{2}\right\rangle =%
\protect\alpha |11\rangle +\protect\sqrt{1-\protect\alpha ^{2}}|00\rangle ,$
respectively.}
\end{figure}

In Fig. 3, 20000 random pure states are generated by Monte Carlo sampling
\cite{new3} and their QSL ratios $\tau /\tau _{QSL}$ and their QSL time
ratios $\tau /\tau _{QSL}$ under memoryless environments versus concurrence
are marked by tiny blue dots. As is clearly displayed in Fig. 3, the lower
bound $\tau /\tau _{QSL}=1$ can always be reached by states $\left\vert \Psi
_{1}\right\rangle =\alpha |01\rangle +\sqrt{1-\alpha ^{2}}|10\rangle $ (dark
red dots in Fig. 3), implying that the concurrence has nothing to do with $%
\tau /\tau _{QSL}$ under this circumstance. In addition, the upper bound is
taken by a subset of states $\left\vert \Psi _{2}\right\rangle =\alpha
|11\rangle +\sqrt{1-\alpha ^{2}}|00\rangle $ (light green dots in Fig. 3),
illustrating that weak entanglement may reduce the QSL time more.

We also note that, if one takes the viewpoint of Scenario 1, the upper
subset of light green dots in Fig. 3 implies that entanglement is able to
accelerate quantum evolution under such a circumstance.

\section{Conclusion}

In summary, we have explored the quantum-speed-limit time for multi-qubit
open systems. For a certain class of initial states, we have demonstrated
that the quantum evolution can also be accelerated in a memoryless
(Markovian) environment. Moreover, we have found that entanglement plays a
subtle role in the speedup of quantum evolution: weak entanglement may be
better for speeding up quantum evolution under certain circumstances.

We have only treated non-correlated environments in this paper. It will also
be of importance and interest to study the QSL time of multi-qubit systems
with the presence of initial correlations among the subsystems of composite
environments \cite{EML}.

\section{\textbf{Acknowledgement}}

This work was supported by the National Natural Science Foundation of China
(Grants No. 11204196 and No. 11074184), the Specialized Research Fund for
the Doctoral Program of Higher Education (Grant No. 20123201120004), and the
Priority Academic Program Development of Jiangsu Higher Education
Institutions.

\end{document}